# Динамика разрывов при цилиндрическом энерговкладе на основе наносекундного разряда

И.А. Знаменская, И.А. Дорощенко
Московский государственный университет имени М. В. Ломоносова, Москва, Россия
E-mail: doroshenko.igor@physics.msu.ru

Экспериментально зарегистрирована сходящаяся вторичная ударная волна, возникающая при реализации протяженного наносекундного сильноточного разряда цилиндрической конфигурации в воздухе. С помощью высокоскоростной теневой съемки (до 300 000 кадров / с) возникающего нестационарного течения визуализированы цилиндрически-симметричные газодинамические разрывы: расходящаяся ударная волна, движущаяся от области плазмы разряда, контактная поверхность, отделяющая ударно-нагретый газ от неравновесной области, возбуждённой плазмой разряда, а также сходящаяся волна сжатия (вторичная ударная волна), движущаяся от контактной поверхности к оси симметрии и фокусирующаяся на ней через 50–60 µs после пробоя с формированием нагретого долгоживущего шнура.

**Ключевые слова:** цилиндрический плазменный канал, наносекундный сильноточный разряд, цилиндрическая ударная волна, высокоскоростная теневая съемка.

В работе изучалась динамика нестационарного разрывного течения, создаваемого наносекундным сильноточным разрядом в форме вертикального цилиндрического канала диаметром $d = 2 - 3$ mm. Разряд инициировался в прямоугольной разрядной камере с кварцевыми окнами между двумя плоскими плазменными электродами, расстояние между которыми составляло $l = 24$ mm. Конденсатор, заряженный до напряжения $U = 25$ kV инициировал при давлениях от 13 kPa до 33 kPa разряд в контрагированном режиме. Экспериментальная установка подробно описана в работах [1, 2, 3]. Визуализация течения осуществлялась с помощью высокоскоростной теневой съемки на параллельных пучках света со скоростью до 300 000 кадров / с. Экспозиция каждого кадра составляла 1 µs. В качестве источника освещения использовался стационарный лазер с длиной волны 532 nm. Также проводилась фото-регистрация свечения разряда при помощи цифрового фотоаппарата.

Ранее было показано, что разряд реализуется в течение 150 – 200 наносекунд (длительность тока) в виде вертикального цилиндрического канала длиной 24 мм и диаметром 2 – 3 мм, и при этом осуществляется режим сверхбыстрого нагрева газа в разрядном канале [1, 2]. На нагрев затрачивается 0.12–0.16 J – значительная часть энергии разряда [2]. В модельной математической постановке такой разряд может рассматриваться как распределенный по цилиндрическому объему взрыв с противодавлением [4, 5, 6]. При таком взрыве на границе газ-плазма формируется первичная ударная волна, контактная поверхность, а также происходит формирование волны разрежения, распространяющейся по газу к центру [4, 5]. Первичная ударная волна ослабевает по мере удаления от центра взрыва за счет потери энергии при расширении и противодавления.

В работе [7] изучалась похожая цилиндрически-симметричная конфигурация разряда и было показано, что на ранней стадии развития течения (t < 100 μs) на границе зоны энерговклада развивается неустойчивость Рэлея–Тейлора. Затем, на более поздних стадиях, формируется неустойчивость, вызванная отклонениями формы канала от симметричной.

Нами были получены теневые фотографии течения, распространяющегося от цилиндрического разрядного канала. На начальных стадиях развития процесса [1], после разряда, регистрировались изображения расходящейся ударной волны в воздухе со средним числом Маха M ≈ 1.5, которая затем покидала область съемки. Также на теневых кадрах присутствует неравновесная область, возбужденная разрядом, ограниченная цилиндрической контактной поверхностью, которая сразу после разряда расширяется до диаметра R = 4 mm и затем останавливается. Волна разрежения, образовавшаяся также как и оба разрыва в начальный момент при распаде разрыва на границе плазменной области, распространяется от границы плазмы к оси симметрии, отражается на ней, а затем догоняет контактную поверхность через 5–10 μs после разряда. Динамика волны разрежения показана в численных расчетах, например, [1, 4]. Результат отражения волны разрежения от контактной поверхности – волна сжатия (слабая ударная волна) (Рис. 1).

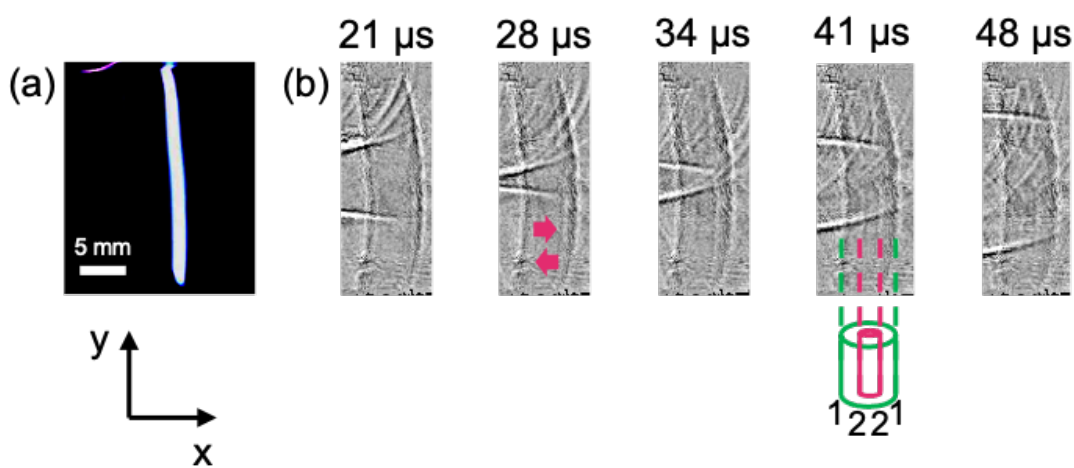

Рис. 1. (a) интегральный снимок разряда; (b) стадии фокусировки вторичной ударной волны: теневые кадры и схема течения после пробоя; видны контактная поверхность (1) и вторичная ударная волна (2), сходящаяся к центру канала. Высота каждого кадра 24 mm.

Детальный анализ процесса выявил внутри области, ограниченной контактной поверхностью, наличие сходящегося цилиндрического разрыва: визуализирована, начиная с 10–12 μs, волна, фокусирующаяся на оси симметрии течения. Наблюдается схождение разрыва к оси с замедлением и, затем, формирование в зоне фокусировки канала радиусом 0.3–0.6 mm. Через 50–80 μs канал теряет устойчивость, его интенсивность на теневых изображениях увеличивается (Рис. 2).

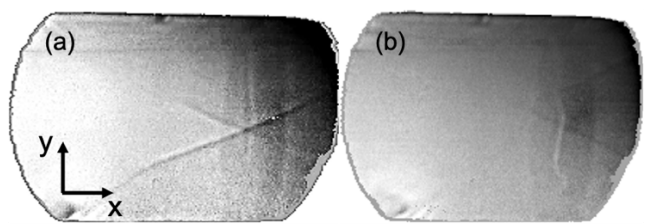

Рис. 2. Теневые кадры канала, сформировавшегося в результате фокусировки сходящейся вторичной ударной волны на оси симметрии – в зоне, ограниченной цилиндрической контактной поверхностью. (a) – стабильная стадия; (b) – неустойчивость канала. Промежуток времени между кадрами 48 µs.

Динамика возникающих цилиндрических разрывов была исследована по сериям полученных теневых кадров и нанесена на x-t диаграммы (Рис. 3). Точки соответствуют усредненным по нескольким экспериментам значениям при давлении в разрядной камере p = 17.6 ± 0.7 kPa. Построены x-t диаграммы движения внешней ударной волны, контактной поверхности и внутренней сходящейся волны. Цилиндрическая ударная волна движется по воздуху от поверхности плазменного канала со средней измеренной скоростью 500 – 520 m/s (M ≈ 1.5). Цилиндрическая контактная поверхность, отделяющая возбужденную разрядом область от потока воздуха за фронтом ударной волны, зафиксирована на теневых снимках начиная с 1–2 µs. После этого она почти не меняет своего положения. Волна сжатия (слабая ударная волна) зарегистрирована с момента 12–14 µs – движется от контактной поверхности к оси симметрии с замедлением по остывающей неоднородной области вплоть до 50–60 µs, со скоростью падающей от 70 m/s до 0. Измеренная скорость существенно меньше соответствующей скорости волн сжатия (звуковой скорости) в совершенном газе. Это связано, очевидно, с неоднородностью параметров остывающей области разряда в радиальном направлении. В результате фокусировки на оси симметрии зарегистрирован цилиндрический нагретый канал радиусом 0.4–0.6 mm. Канал остается стабильным в течение десятков микросекунд, а затем начинает изгибаться, теряя свою устойчивость.

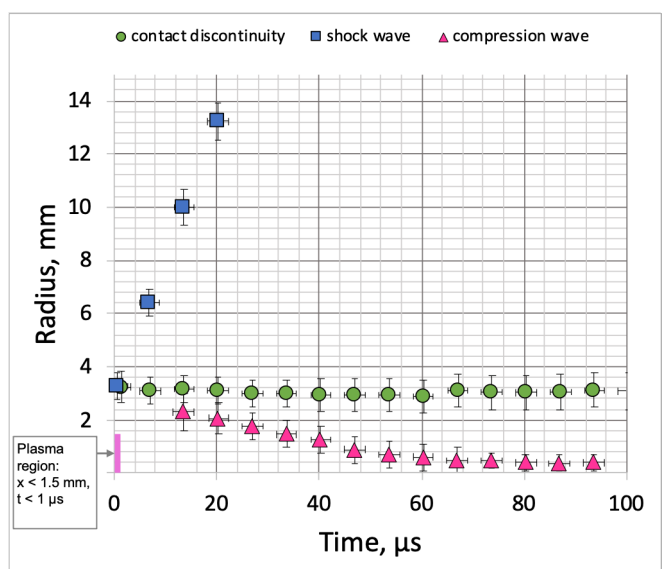

Рис. 3. x-t диаграмма движения разрывов после разряда по результатам высокоскоростной теневой съемки.

Стоит отметить, что эффект возникновения радиального движения газа к оси симметрии для рассматриваемого течения был также ранее зафиксирован при визуализации полей скоростей методом PIV (англ. Particle Image Velocimetry) [2].

Задача о фокусировке сильной цилиндрической или сферической ударной волны в идеальном газе была рассмотрена независимо друг от друга Г. Гудерлеем в 1942 году, Л.Д. Ландау в 1944 г. и К.П. Станюковичем в 1945 г. Современное изложение задачи приведено в книге [8]. Численное решение задачи приведено в работах [9, 10]. Фокусировка ударной волны может быть использована, например, для управляемой детонации газовых пузырей различной формы [11]. В работах [12, 13, 14] изучалась процессы фокусировки и кумуляции сходящейся ударной волны при ее инициировании на периферии цилиндрического объема.

В данной работе изучается возникновение и фокусировка вторичной ударной волны при распределенном по цилиндрическому объему взрыве с противодавлением. В результате фокусировки не происходит отражения волны от оси симметрии, а формируется вертикальный канал вдоль нее.

Данная задача является весьма сложной для анализа, так как значения параметров газа (плотности, скорости звука) в области пробоя, по которой движется вторичная сходящаяся ударная волна, неизвестны, они могут изменяться во времени и пространстве. Исследуемое в данной работе течение характеризуется эффектом «перерасширения», результатом которого является остановка контактного разрыва (в отличие от взрыва в среде без противодавления) и формирование за ним ударной волны, распространяющейся от контактного разрыва к оси симметрии и затем фокусирующейся с образованием узкого вертикального канала. Рассматриваемая задача является цилиндрически симметричной.

Таким образом, в данной работе зафиксирован экспериментально эффект возникновения и фокусировки вторичной сходящейся ударной волны для задачи взрыва, вызванного пробоем импульсного протяженного сильноточного разряда цилиндрической геометрии (длина 24 мм диаметр 2–3 мм). Получены теневые кадры газодинамического процесса, исследована динамика возникающих внешней ударной волны, контактной поверхности и внутренней (вторичной) волны, фокусирующейся на оси симметрии течения и формирующей там область нагрева. Показано что скорость движущегося к оси от контактной поверхности вторичного разрыва падает от 70 m/s до 0.